# Ranking Journals Using Altmetrics[1]

Tamar V. Loach[1,2] and Tim S Evans[2]

[1]Digital Science, 2 Trematon Walk, Wharfdale Road, London, N1 9FN (U.K.)
[2]Imperial College London, Centre for Complexity Science, South Kensington Campus, London SW7 2AZ (U.K.)

**Abstract**

The rank of a journal based on simple citation information is a popular measure. The simplicity and availability of rankings such as Impact Factor, Eigenfactor and SciMago Journal Rank based on trusted commercial sources ensures their widespread use for many important tasks despite the well-known limitations of such rankings.

In this paper we look at an alternative approach based on information on papers from social and mainstream media sources. Our data comes from altmetric.com who identify mentions of individual academic papers in sources such as Twitter, Facebook, blogs and news outlets.

We consider several different methods to produce a ranking of journals from such data. We show that most (but not all) schemes produce results, which are roughly similar, suggesting that there is a basic consistency between social media based approaches and traditional citation based methods. Most ranking schemes applied to one data set produce relatively little variation and we suggest this provides a measure of the uncertainty in any journal rating. The differences we find between data sources also shows they are capturing different aspects of journal impact. We conclude a small number of such ratings will provide the best information on journal impact.

**The background and purpose of the study**

Journal metrics, such as the Thomson Reuters Journal Impact Factor, were originally developed in response to a publisher need to demonstrate the academic attention accorded to research journals. Over the intervening 50 years since Garfield's work in the field, the Impact Factor and other metrics, such as Eigenfactor (Bergstrom 2007), have been used and misused in a variety of contexts in academia. An oft-discussed perception is that a journal-level metric is a good proxy for the quality of the articles contained in a journal.

In the evaluation and bibliometrics communities citation counting is generally understood not to be an appropriate proxy for quality but rather a measure of attention. The type of attention being measured in this case is quite specific and has particular properties. What is being measured is the attention to a paper of peers in related fields. The bar for registration of this attention is relatively high – the researcher or researchers making the citation must deem the target article to be of sufficient value that they include a citation in a work of their own that in turn is deemed publishable (e.g. see Archambault & Lariviére 2009 and references therein). The timescale associated with citations is also long – typically being limited by the review and publication process associated with particular fields. Additionally, it is accepted that journal-level metrics say little regarding the merit of particular articles in the journal since journal-level metrics are often calculated based on thousands of articles and are often biased by the performance of the tails of the distribution of citations. These realisations have led to the recent growth in popularity of article-level metrics or altmetrics.

Altmetrics have broadened the range of types of attention that we can measure and track for scholarly articles. Mostly based in social and traditional media citations, the altmetric landscape is one that is constantly changing with the introduction of different data sources all the time. While, one the one hand, altmetrics suffer from all the unevenness of traditional citations, they occur over different timescales, which provides us with a more nuanced view of the lifecycle of a scholarly work. Aggregating alternative metrics at a journal level will complement Journal Impact Factor, giving us new insights into different facets of attention.

---

Traditional citation-based metrics are difficult to calculate since they are based on the bibliometric journal databases, such as Thomson Reuters' Web of Science. Conversely, Altmetrics are conglomerates of disparate sources of references to research output derived from non-traditional sources, primarily modern electronic sources characterised by fast response times (see Bornmann, 2014 for a recent overview). The lack of any systematic peer review is another characteristic of most altmetric data. The open and electronic nature of much altmetric data offers the prospect of alternative paper and journal metrics, which may be more accessible to stakeholders. The rapid response of such data to innovations suggests such metrics might offer improvements over metrics based on slower traditional sources.

This paper considers a number of approaches to the aggregation of altmetric data in order to create a robust journal-level metric that complements the existing citation-based metrics already in use across the academic community. The aim is not to create a contender for a single metric to quantify journal output but instead to create a useful measure that gives "the user" a sense of the non-citation attention that a journal attracts in the same way that Journal Impact Factor, Eigenfactor and other related metrics give this sense for citation attention.

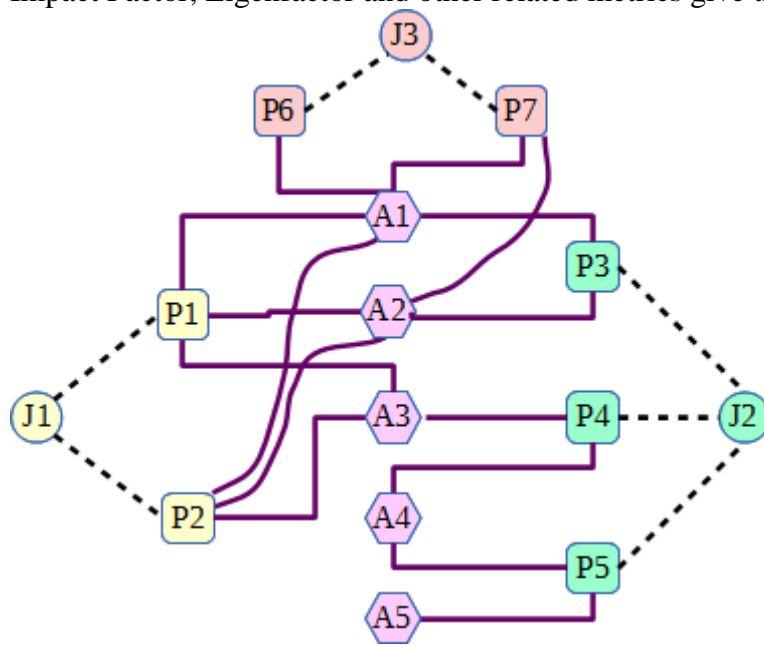

Figure 1. The relationships recorded in our altmetric.com data. The raw data illustrated here contains fifteen "mentions" (solid lines) by five "authors" (hexagons A1 to A5) of seven papers (squares P1 to P7). We also know the journal (circles), which published a paper (dashed lines).

**Data Sources**

In this paper we use the 2013 IF (Impact Factor) and EF (Eigenfactor) as examples of traditional sources of journal ratings. Our altmetric data comes from 20 months of data from altmetric.com, a commercial company. For each mention about a paper we had the journal in which it was published, the source (twitter, Facebook, etc.) and the account (here termed an 'author'), as shown in Figure 1. In our case, a 'paper' has to be an article coming from a known journal. A single 'author' for us is a single account (e.g. one twitter account) or a single source (a news outlet such as a newspaper). In some cases several different authors may be responsible for one site or one author could provide information to many different sites or accounts (a twitter account, a facebook account, a blog, etc) but in our data such an author appears as many distinct authors.

**Methods**

The simplest type of journal altmetric is one based on basic counts where each mention of a paper in a journal adds one to that journal's count. We collected counts for social media '*sbc*', non-social media '*nsbc*' (e.g. downloads) and combined scores '*bc*' (for blind count i.e. with no weighting for different sources). We also obtained the current journal rating produced by

altmetric.com (denoted '*ca*'), which is a weighted count rating in which different sources are given different weights (blogs and news sources get highest weighting).

*Network Definitions*

A criticism of simple count based methods, such as Impact Factor or our altmetric counts discussed above, is that some citations or some altmetric authors are more important than others. Eigenfactor is an illustration of a response to these criticisms in the realm of traditional data (Bergstrom 2007), as it uses a network based view to arrive at a PageRank style measure. We will also turn to a network-based view in order to look at a wide range of measures, which probe the relationships between journals on a much larger scale.

There are many possible network representations of our data. In this paper we will focus only on networks in which the nodes represent journals. The central idea in our construction of the relationship between two journals is that we only want to consider activity from authors who mention both journals because only these authors are making an implicit comparison between journals. The activity of each author is used to define their own "field of interest" in a self-consistent manner and so the activity of authors is used to make comparisons between journals in the same field as defined by each author's interests. This ensures that at a fundamental level we avoid the much discussed problem of making comparisons between papers or journals from different fields. An author only interested in medical issues will only contribute to the evaluation of Nature, Science and so forth in terms of their interest in these multidisciplinary journals relative to Cell or other specialised journals.

A useful analogy here is that each journal is a team and an author who mentions articles published in two journals represents one game between these journals – our pairwise comparison. The score in each game is the number of mentions so in comparing two journals *j* and *l*, the score for journal *j* from the game represented by author *a* is recoded as the entry $J_{ja}$ in a rectangular matrix. In Figure 1 the game between J1 and J2 represented by author A2 has the result 2-1, a 'win' for journal 1 over journal J2 suggesting that we should rate journal J1 more highly than journal J2 given the activity of this one author.

We shall consider three different ways of quantifying the journal relationships, the network edges. Our first approach gives us an adjacency matrix *S* where the entry $S_{jl}$ gives the weight of the edge from journal *j* to journal *l*, and this is given by $S_{jl} = \frac{1}{|A_{jl}|} \sum_{a \in A_{jl}} J_{ja}$ where $A_{jl} = \{a | J_{ja} > 0, J_{la} > 0 \}$. Here *j* and *l* represent different journals and *a* is one author. $J_{ja}$ is a matrix, which is equal to the number of papers mentioned by author *a* which were published in journal *j*. The expression for $S_{jl}$ is counting the number of times papers published in journal *j* are mentioned by authors who also mention papers in journal *l*, with the total normalised by the number of such authors. Note that this defines a sparse, weighted and directed network. In our conventions if journal *j* is better than journal *l* we will have $S_{jl} > S_{lj}$.

Our second definition gives us an adjacency matrix *P* where $P_{jl} = \frac{1}{|A_{jl}|} \sum_{a \in A_{jl}} \theta(J_{ja} - J_{la})$. Here $\theta(x) = 1$ if $x > 0$ otherwise this function gives 0. This definition counts how many authors mention more papers in journal *j* than they do papers in journal *l*., normalising again by the number of authors who are able to make this pairwise comparison. Again $P_{jl} > P_{lj}$ if journal *j* is better than journal *l*.

Finally we define an adjacency matrix *Q* where $Q_{jl} = \frac{1}{|A_{jl}|} \sum_{a \in A_{jl}} \Theta(J_{ja} - J_{la})$. Here $\Theta(x) = 1$ if $x > 0$, $\Theta(0) = 0.5$ while for negative values this function gives 0. This definition counts how many authors mention more papers in journal *j* than they do papers in journal *l* but when this is balanced gives an equal weighting to both side. This definition has the useful property that $Q_{jl} + Q_{lj} = 1$ (not generally true for matrix *P*).

*Network Measures*

Once we have our network with journals as nodes, we need to find ways to use this structure to define which nodes are the most important. Measures which quantify the importance of a node are known as centrality measures in social network analysis. Unfortunately, many standard measures do not take into account the weights or directions of edges, both of which carry crucial information in our case. We used two well-known network centrality measures to illustrate our approach: PageRank and HITS (e.g. see Langville & Meyer 2012). Both may be cast as eigenvector problems and there are fast algorithms for large networks which are readily available. We apply these two methods to all three networks, giving six different ratings e.g. 'qpr' indicates a PageRank rating derived from a *Q* matrix while 'ph' indicates a HITs rating derived using a *P* matrix.

We also tried a different type of measure known as Points Spread Rating (denoted 'psr') (p117-120, Langville & Meyer 2012) where the rating $r_j$ for journal *j* is $r_j = \sum_l (S_{jl} - S_{lj})/n_j$, (similarly for the *P* and *Q* matrices) and $n_j$ is the number of journals. This expression ensures that the differences *($r_j$-$r_l$)* in the rating of any two journals *j* and *l* are as close as possible to the actual differences in the number of average mentions of papers.

*Comparing Ratings*

Once we have obtained different ratings, the final task is to make a comparison. The simplest approach is to make a qualitative comparison of the top ranked journals in each case. For a more quantitative approach we used standard methods of multivariate statistics. First we found a correlation matrix whose entries express the similarity of two rating methods: the Pearson correlation matrix based on the numerical values of the ratings obtained, Spearman's matrix which based on the ranking of journals, and finally Kendall's tau. These were analysed using principle component analysis or hierarchical clustering methods.

**Findings**

In terms of the altmetric data we found typical fat-tailed distributions, both for the number of mentions of a paper from different sources and in terms of the number of mentions put out by a single author. Some sources, such as twitter, are significantly larger than others.

When comparing different journal rating schemes, some results were found only with Spearman and Kendall tau correlation measures (which are based on the ranks of journals). The Pearson measure (based on actual rating values gave slightly different results in some cases. However in most cases there good agreement. Some typical results are shown in Figure 2 and numbers for ranking schemes in the following text refer to the labels in Figure 2.

The variation between different rating schemes for the same altmetric data source gives relatively little variation, roughly on the same scale as the difference we find between IF and EF. The four different methods shown for ratings based on Facebook mentions (6,12,16,19) are a typical example. Clearly our Points Spread Rating scheme (psr, 21,22,23) and our simple counts of non-social media mentions (nsbc, 6) produces outliers.

Some sources, such as Facebook and News, were also noticeably different from IF and EF, but the difference was much smaller than that found with the psr rating. One source, which gave ratings well correlated with IF and EF was blogs (8, 11, 15, 18).

Likewise, most of our simple count based ratings were just as close to IF (3) or EF (5) as these two rating schemes were to each other. This includes our unweighted count of all mentions (bc, 1), the number of times papers are mentioned (pc, 7), counts of just social media mentions (sbc, 14), and in particular the more sophisticated weighted journal ranking produced by altmetric.com (ca, 2).

Most of our work focused on statistics for the whole collection. A look at the top journals, see

Table 1, confirmed that at an individual level our new altmetric network ratings were giving sensible results, but with variations which indicate the uncertainty in such rankings.

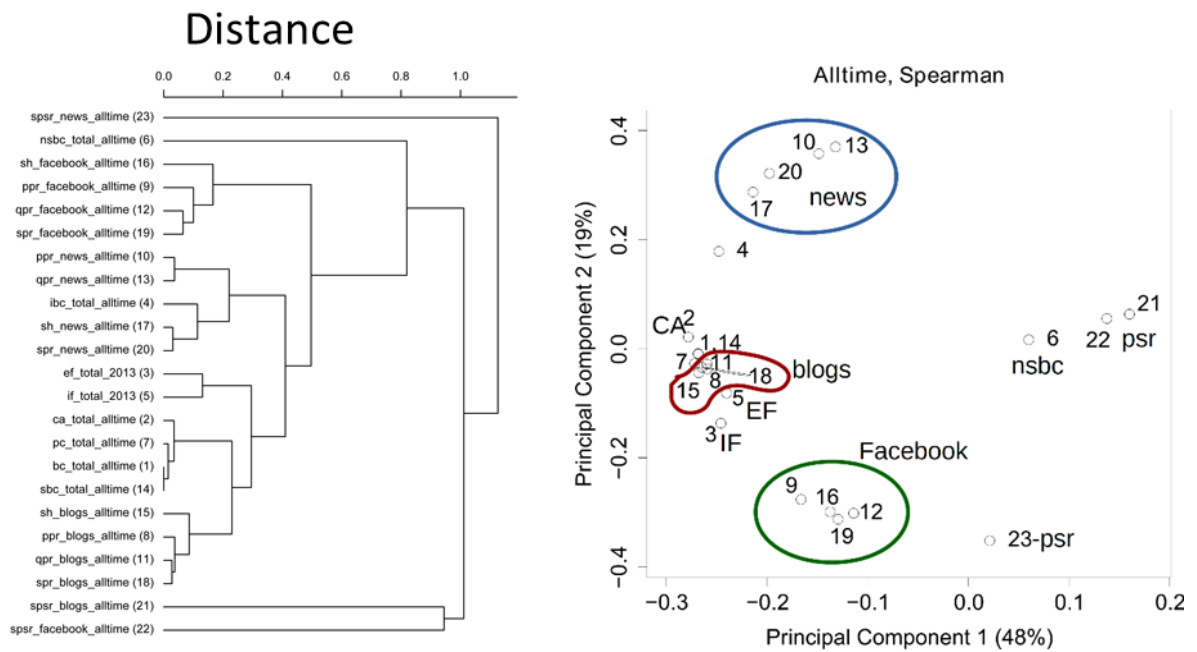

**Figure 2. A comparison of some of the different ranking schemes using a Spearman correlation matrix. On the left a dendrogram and on the right a scatter plot using the first two principle components of PCA. For clarity, only a limited subset of our ratings were used in these plots.**

**Discussion**

Given our differences between ranking based comparisons (Spearman and Kendall Tau) and results based on Pearson correlation matrices, this suggests that ratings are dominated by the measurement of the few journals, which have most of the mentions (fat tails). This is one reason we favour Spearman correlation matrices in Figure 2 and would suggest this makes sense in most journal ranking contexts.

Our Points Spread Rating scheme (psr, 21, 22, 23) seems to be reflecting very different patterns in the data from those found using other approaches. Given that the other approaches include Impact Factor, widely accepted as a measure of journal attention, we think it is hard to see a role for PSR to rank journals. Likewise, the simple blind counts of non-social media mentions (nsbc, 6) does not appear to be useful.

The remaining different altmetric sources and rating methods do show enough similarity to suggest that they are all an acceptable measure of journal importance. At the same time there are some interesting differences indicating that our altmetric based schemes are capturing different features of the impact of journals. At the very least this diversity will indicate the level of uncertainty in rating schemes. Two possible reasons for the close correlation of blogs and IF are as follows. Perhaps papers in high IF journals are of intrinsic interest to blog writers. Alternatively blog authors may read a limited number of journals but these tend to be those with high IF. Probably both factors are important, each reinforcing the other to produce the strong correlation we find.

Another interesting feature is that most of our simple count based ratings, which are not normalised by the number of articles per journal, are also well correlated with IF (3) which does use normalised counts. This can be explained if there is a correlation between the number of papers in a journal and its impact, something we can see in of count of number of papers (pc, 7). We will be looking at normalised altmetric counts in the future but it appears normalisation may not be essential. In particular, we note the altmetric.com journal rating (ca, 2) is well correlated and so provides a good handle on the impact of journals.

**Table 1 Top ten journals based on various network based altmetric measures.**

| Rank | Q, HITS, Blogs | Q, HITS, News | S, PageRank, Google+ |
|---|---|---|---|
| 1 | Nature | Nature | Nature |
| 2 | PNAS | PNAS | PLoS ONE |
| 3 | Science | PLoS ONE | Science |
| 4 | PLoS ONE | Science | PNAS |
| 5 | New England J. of Med. | New England J. of Med. | New England J. of Med. |
| 6 | British Medical J.-C.R.Ed. | British Medical J.-C.R.Ed. | British Medical J.-C.R.Ed. |
| 7 | The Lancet (British Ed.) | Nature Communications | Scientific Reports |
| 8 | JAMA | JAMA | JAMA |
| 9 | Proc. Royal Soc. B: | The Lancet (British Ed.) | The Lancet (British Ed.) |
| 10 | Current Biology | Pediatrics | PLoS Biology |

The fact that we tried many different rating methods and that (with the exception of psr based measures) they showed variations on scales no bigger than those found between IF and EF, suggests that no one method is optimal in any sense. However we can use such a suite of metrics to get a handle on the uncertainty associated with any measure. This would be of great utility for users and a contrast to the three decimal point 'accuracy' associated with IF results.

**Conclusions**

We have shown how to use altmetric data to provide a reasonable journal ranking. Most types of altmetric data appear to give useful information in the sense that the correlation with IF is acceptable. At the same time altmetric data can be sufficiently different that it might reflect different types of impact. Our results suggest that different rating methods can provide a measure of the uncertainty of any journal ranking. Confirming these patterns over longer periods and producing a better understanding of the social reasons for the patterns we have found are future directions for our work. It would also be interesting to compare our results with journal attention measures derived from journal usage patterns, see for example Bollen et al 2009, an aspect not included in our data.

**Acknowledgments**

We would like to thank Euan Adie (altmetric.com) for providing us with the altmetric data, and for useful discussions along with Jonathan Adams and Daniel Hook (Digital Science).